\documentclass[
  journal=largetwo,
  manuscript=article-type,
  year=2025,
  volume=37,
]{cup-journal}

\usepackage{amsmath}
\usepackage{xspace}
\usepackage[nopatch]{microtype}
\usepackage{booktabs}
\usepackage{graphicx}
\usepackage{makecell}
\usepackage{amssymb}
\usepackage{siunitx}
\usepackage{txfonts}
\usepackage{soul}
\usepackage{hyperref}
\hypersetup{
    breaklinks=true,
    colorlinks=true,
    linkcolor=blue,
    citecolor=blue,
    urlcolor=blue,
    filecolor=magenta,
    pdftitle={Overleaf Example},
    pdfpagemode=FullScreen,
}

\title{A Search for Planet Nine with IRAS and AKARI Data}

\author{Terry Long Phan}
\affiliation{Institute of Astronomy, National Tsing Hua University, 101, Section 2, Kuang-Fu Road, Hsinchu, 30013, Taiwan}
\email[Terry Long Phan]{terryphan224@gapp.nthu.edu.tw}

\author{Tomotsugu Goto}
\affiliation{Institute of Astronomy, National Tsing Hua University, 101, Section 2, Kuang-Fu Road, Hsinchu, 30013, Taiwan}
\alsoaffiliation{Department of Physics, National Tsing Hua University, 101, Section 2, Kuang-Fu Road, Hsinchu, 30013, Taiwan}

\author{Issei Yamamura}
\affiliation{Institute of Space and Astronautical Science, Japan Aerospace Exploration Agency, 3-1-1 Yoshinodai, Chuo-ku, Sagamihara, Kanagawa 252-5210, Japan}

\author{Takao Nakagawa}
\affiliation{Institute of Space and Astronautical Science, Japan Aerospace Exploration Agency, 3-1-1 Yoshinodai, Chuo-ku, Sagamihara, Kanagawa 252-5210, Japan}
\alsoaffiliation{Advanced Research Laboratories, Tokyo City University, 1-28-1 Tamadutsumi, Setagaya, Tokyo 158-8557, Japan}

\author{Amos Y.-A. Chen}
\affiliation{Department of Physics, National Tsing Hua University, 101, Section 2, Kuang-Fu Road, Hsinchu, 30013, Taiwan}

\author{Cossas K.-W. Wu}
\affiliation{Institute of Astronomy, National Tsing Hua University, 101, Section 2, Kuang-Fu Road, Hsinchu, 30013, Taiwan}

\author{Tetsuya Hashimoto}
\affiliation{Department of Physics, National Chung Hsing University, 145, Xingda Road, Taichung, 40227, Taiwan}

\author{Simon C.-C. Ho}
\affiliation{Research School of Astronomy and Astrophysics, The Australian National University, Canberra, ACT 2611, Australia}
\alsoaffiliation{Centre for Astrophysics and Supercomputing, Swinburne University of Technology, Hawthorn, VIC 3122, Australia}
\alsoaffiliation{OzGrav: The Australian Research Council Centre of Excellence for Gravitational Wave Discovery, Hawthorn, VIC 3122, Australia}
\alsoaffiliation{ASTRO3D: The Australian Research Council Centre of Excellence for All-sky Astrophysics in 3D, ACT 2611, Australia}

\author{Seong Jin Kim}
\affiliation{Institute of Astronomy, National Tsing Hua University, 101, Section 2, Kuang-Fu Road, Hsinchu, 30013, Taiwan}


\begin{document}

\begin{abstract}
The outer solar system is theoretically predicted to harbour an undiscovered planet, often referred to as Planet Nine. Simulations suggest that its gravitational influence could explain the unusual clustering of minor bodies in the Kuiper Belt. However, no observational evidence for Planet Nine has been found so far, as its predicted orbit lies far beyond Neptune, where it reflects only a faint amount of Sunlight. This work aims to find Planet Nine candidates by taking advantage of two far-infrared all-sky surveys, which are IRAS and \textit{AKARI}. The epochs of these two surveys were separated by 23 years, which is large enough to detect Planet Nine's $\sim3'$/year orbital motion. We use a dedicated AKARI Far-Infrared point source list for the purpose of our Planet Nine search --- AKARI-FIS Monthly Unconfirmed Source List (AKARI-MUSL), which includes sources detected repeatedly only in hours timescale, but not after months. AKARI-MUSL is more advantageous than the AKARI Bright Source Catalogue (AKARI-BSC) for detecting moving and faint objects like Planet Nine with a twice-deeper flux detection limit. We search for objects that moved slowly between IRAS and \textit{AKARI} detections given in the catalogues. First, we estimated the expected flux and orbital motion of Planet Nine by assuming its mass, distance, and effective temperature to ensure it can be detected by IRAS and \textit{AKARI}, then applied the positional and flux selection criteria to narrow down the number of sources from the catalogues. Next, we produced all possible candidate pairs including one IRAS source and one \textit{AKARI} source whose angular separations were limited between $42'$ and $69.6'$, corresponding to the heliocentric distance range of 500 -- 700 AU and the mass range of 7 -- 17M$_{\oplus}$. There are 13 candidate pairs obtained after the selection criteria. After image inspection, we found one good candidate, of which the IRAS source is absent from the same coordinate in the \textit{AKARI} image after 23 years and vice versa. However, \textit{AKARI} and IRAS detections are not enough to determine the full orbit of this candidate. This issue leads to the need for follow-up observations, which will determine the Keplerian motion of our Planet Nine candidate.
\end{abstract}

\section{Introduction}
The discoveries of Sedna \citep{Brown2004} and Sedna-like bodies \citep[e.g.,][]{trujillo2014sedna, Sheppard_2019} revealed their unusual orbital properties such as high eccentricity and inclination. In addition, the simulation work of \citet{Batygin2016} showed that the orbits of various distant Kuiper Belt Objects (KBOs) exhibited strong clustering in both arguments of perihelion and physical space. The probability that this peculiar clustering happens by random processes is only 0.007\%, which is almost impossible unless it is related to another unknown dynamical process.

While \citet{Brown_2017} implied that the observational biases might cause the observed clustering of distant KBOs, \citet{Brown2019} developed a method to quantify the observational biases and concluded that the chance for the observed random clustering is only 0.2\%. Finding the elusive mechanism behind the orbital clustering of KBOs has become an attractive topic because it challenges the contemporary understanding of the outer solar system.  \citet{Batygin2016} proposed that a massive and distant planet, also known as Planet Nine, can maintain the orbital alignment of those KBOs with mass $\geq$ 10M$_{\oplus}$ and semi-major axis of 700 AU. Since this is an undiscovered planet, several names have been proposed to call it. In 1906, Percival Lowell started to search for a hypothetical planet beyond Neptune's orbit and called it Planet X\footnote{The letter X represents the unknown status of the planet, not the Roman numeral for 10.}. The next name, Planet Nine or Planet 9, was proposed after the demotion of Pluto from a planet to a dwarf planet in 2006 and has been widely used in recent studies. \citet{Iorio_2017} proposed the name Telisto\footnote{The name Telisto should not be confused with Telesto, one of the natural satellites of Saturn.} because while the official name of the object is supposed to be changed in the future according to the classification issues, its great distance is an unchanged feature. The speculation of Planet Nine's orbit and characteristics, such as mass and distance, was followed by various simulation works \citep[e.g.,][]{Brown_2016, Batygin_2016, Millholland2017, Brown2021}, in which the N-body simulations and Markov Chain Monte Carlo analysis were employed to estimate parameters of Planet Nine. Furthermore, the perturbations caused by a distant point-like body on a two-body system like the Sun and Earth were analytically calculated in \citet{Iorio_2024}. Attempts to constrain the possible location of Planet Nine have been made recently through different released observational data \citep[e.g.,][]{Iorio_2013, Fienga_2016, Holman_2016, Fienga_2020} and its perturbations on other Solar System bodies \citep[e.g.,][]{Iorio_2017, Nabiyev_2022, Gomes_2023}. In addition, the Planet Nine-inclusive model was compared to the Planet Nine-free model to assess the influence on the low-inclination, Neptune-Crossing TNOs with and without the presence of Planet Nine \citep{Batygin_2024}, where the Planet Nine-free scenario was rejected at a $\sim$ 5$\sigma$ confidence-level. An alternative explanation suggested that our solar system captured one of the primordial black holes (PBHs) with 5M$_{\oplus}$ $\leq$ M$_{BH}$ $\leq$ 10M$_{\oplus}$ \citep{PhysRevLett.125.051103}, which can explain why Trans-Neptunian Objects (TNOs) move perpendicularly to the planetary plane and why the orbits of extreme TNOs exhibit an unexpected clustering. Although the orbital clustering of KBOs was statistically demonstrated not to occur by chance, the existence of Planet Nine has been under debate so far. One previous study using an observed sample of large semi-major axis TNOs from the Outer Solar System Origins Survey \citep{Shankman_2017} concluded that the evidence for a super-Earth or larger planet causing the orbital clustering of those TNOs is in doubt. However, the scenario of a larger-than-dwarf-scale planet is still possible. Another study on 14 extreme TNOs discovered by the Dark Energy Survey, the Outer Solar System Origins Survey, and the survey of Sheppard and Trujillo \citep{Napier_2021} obtained a joint probability between 17\% and 94\% for the uniform distribution of the extreme TNOs, which is different from the value of 0.2\% in \cite{Brown2019}. While the results from \citet{Napier_2021} indicated that there is no evidence for the angular clustering, they could not explicitly rule out Planet Nine with current data.

Many efforts in searching for Planet Nine have been made recently by utilising optical surveys such as Zwicky Transient Facility \citep[ZTF,][]{Brown2022}, Dark Energy Survey \citep[DES,][]{Belyakov2022}, and Pan-STARRS1 \citep[PS1,][]{Brown2024}. However, they only ruled out percentages of the Planet Nine parameters predicted by \citet{Brown2021} without successfully constraining any Planet Nine candidates. One possible explanation for this problem is that the predicted semi-major axis of Planet Nine's orbit is 700 AU \citep{Batygin2016}, which is roughly 23 times larger than that of Neptune's orbit ($\sim$ 30 AU). Such a distant location makes Planet Nine too dim to observe from Earth in optical wavelengths. When increasing the current heliocentric distance $d$ of Planet Nine, the Sunlight reflected by Planet Nine decreases as a function of $d^{-4}$, while the thermal radiation in infrared wavelengths of this planet only weakens as a function of $d^{-2}$. As a result, Planet Nine in infrared wavelengths is expected to be much brighter than in optical wavelengths. On the other hand, when infrared light travels from the outer solar system to the Earth, it is absorbed by the Earth's atmosphere. In this case, infrared surveys using space telescopes instead of ground-based telescopes are able to improve signals from distant, faint sources. The Backyard Worlds: Planet 9 Citizen Science Project\footnote{\url{https://www.zooniverse.org/projects/marckuchner/backyard-worlds-planet-9}} was conducted by using data from Wide-field Infrared Survey Explorer (WISE) mission. \citet{Rowan2021} searched for Planet Nine in Infrared Astronomical Satellite (IRAS) data and found one Planet Nine candidate with a distance of 225 
$\pm$ 15 AU and a mass of 3 -- 5M$_{\oplus}$. Another far-infrared search was implemented by \citet[][hereafter SS22]{Sedgwick2022} with a distance range of 700 -- 8000 AU. Their idea was to find giant planet candidate pairs, including one IRAS detection and one \textit{AKARI} detection. They found 535 potential candidate pairs after the Spectral Energy Distribution (SED) fitting. However, all of their candidates are located in the galactic cirrus region, which affects the reliability of flux measurements. Several recent studies proposed and evaluated the prospects of future planetary and deep-space missions for the Planet Nine search, including a dedicated mission to measure modifications of gravity out to 100 AU \citep{Buscaino_2015}, the Uranus Orbiter and Probe mission \citep{Bucko_2023}, and the Elliptical Uranian Relativity Orbiter mission \citep{Iorio_2023}.

Similar to SS22, our objective in this study is to take advantage of the data of two far-infrared all-sky surveys, which are IRAS and \textit{AKARI}, and then find candidates for the mysterious planet. However, the key difference between our work and SS22 is distinguished by the AKARI catalogues. Due to Earth's orbital motion around the Sun, Planet Nine's parallax is expected to shift $\sim 10' - 15'$ every 6 months, corresponding to the heliocentric distance range of 500 -- 700 AU. Therefore, it will not be observed at the same location after this period. We use \textit{AKARI} sources without monthly confirmation instead of the co-added catalogue like AKARI-BSC Version 2, which was used in SS22. In other words, since the position of Planet Nine is affected by the parallax motion and the orbital motion, we are only able to detect Planet Nine at the same location over several successive scans in a day, but not after half a year. This paper is organized as follows: Section \ref{sec:data} presents a brief introduction to the prospect and the power of two surveys as well as the data sets we use in this work, Section \ref{sec:method} includes our method to estimate the expected orbital motion and flux of Planet Nine, and a set of criteria for positional and flux selections. We show our results in Section \ref{sec:results}, discussions in Section \ref{sec:discussion}, and summarise the outcomes of this research in Section \ref{sec:sum}.

\section{Far-Infrared All-Sky Surveys}
\label{sec:data}
Although there are various predictions for the possible sky region or the orbital structure of Planet Nine in the outer solar system \citep[e.g.,][]{Brown_2016, Batygin_2016, Millholland2017, Brown2021}, its precise location remains unconfirmed. That is the main reason we need all-sky surveys to conduct a large-area search. On the other hand, since Planet Nine could be a few hundred astronomical units away from Earth, the motion of Planet Nine in a short time scale of several months or a year is not significant enough to be detected. That means it requires at least two all-sky surveys with a sufficient epoch separation. Therefore, in this paper, we chose two all-sky surveys with a 23-year epoch separation. The characteristics of these two surveys are described below and summarised in Table \ref{tab:data}.

\subsection{IRAS}
The Infrared Astronomical Satellite \citep[IRAS,][]{neugebauer1984infrared} was launched in 1983 to carry out the all-sky survey. This mission covers four infrared wavelength bands centred at 12 $\mu$m, 25 $\mu$m, 60 $\mu$m and 100 $\mu$m. There are two main catalogues released by the IRAS team: the Point Source Catalogue (IRAS-PSC) includes over 245,000 sources, and the Faint Source Catalogue (IRAS-FSC) includes over 173,000 sources. The flux detection limit of the IRAS depends on the catalogue type and the wavelength. To be specific, at 100 $\mu$m, both IRAS-PSC and IRAS-FSC have a detection limit of 1.0 Jy. At 60 $\mu$m, however, the detection limit of IRAS-PSC is 0.6 Jy, whereas that of IRAS-FSC can reach 0.2 Jy. 

In addition, there are also two other catalogues, including rejected sources of two main catalogues: the Point Source Catalogue Rejects (IRAS-PSCR) and the Faint Source Catalogue Rejects (IRAS-FSCR). The reasons for rejection could be the failure to achieve the minimum criterion of two-hour confirmation, or sources located in confused regions of the sky or were only detected in a single band. As a result, when we look at the co-added images of sources in these two rejected catalogues, they might not be shown as real physical sources. Despite the rejection, there is still a possibility that Planet Nine would be detected in one of these catalogues. 

IRAS team matched their sources with other catalogues to make a source identification, indicated by the number of positional associations ($nid$). This parameter has integer values ranging from 1 to 4. Each value of $nid$ represents a type of catalogue associated with an IRAS source, such as extragalactic catalogues ($nid$ = 1), stellar catalogues ($nid$ = 2), other catalogues ($nid$ = 3), and multiple types of catalogues ($nid$ = 4). If sources are not found in any type of catalogue listed above, $nid$ = 0. The total numbers of identified and unidentified sources are shown in Table \ref{tab:data}.

\subsection{AKARI}
Another far-infrared all-sky survey was conducted by the Japanese infrared astronomical satellite \textit{AKARI} in 2006 -- 2007 \citep{Murakami2007}. \textit{AKARI} observed the whole sky with four far-infrared bands centered at 65 $\mu$m (N60), 90 $\mu$m (WIDE-S), 140 $\mu$m (WIDE-L), and 160 $\mu$m (N160). The AKARI-FIS Bright Source Catalogue (Version 1 and Version 2) has been published for general astronomical research. In this work, however, one of the authors (IY) has created a custom-made source list, "AKARI-FIS Monthly Unconfirmed Source List" (AKARI-MUSL), to search for Planet Nine candidates. The AKARI-MUSL is based on the same internal dataset of point-source-like signals detected in each scan during the survey used for the AKARI-BSC, but with different source-selection criteria. \textit{AKARI} survey was carried out over 16 months from May 2006 to August 2007. A position of the sky was observed in a few to several successive scans (the number of scans depends on the Ecliptic latitude), then repeated a half year later. 

\begin{figure}
   \centering
   \includegraphics[width=\columnwidth]{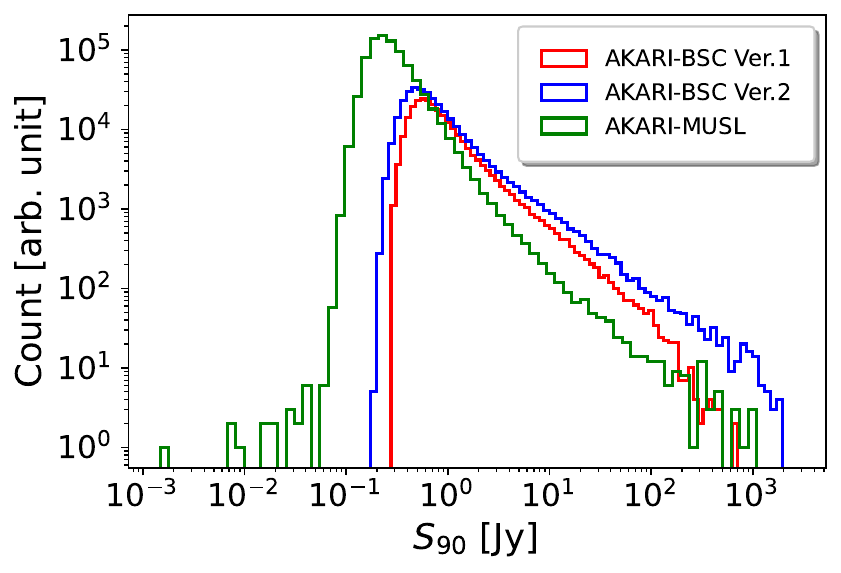}
      \caption{Comparison between log $N$ -- log $S$ plots of three different AKARI catalogues in 90 $\mu$m: AKARI-BSC Version 1 (red), AKARI-BSC Version 2 (blue), and AKARI-MUSL (green).
              }
         \label{fig:detection_limit}
\end{figure}

In AKARI-BSC, a "real source" is defined as one that is detected in $\geq$ 75\% of the total number of scans at the same position (e.g., 10 detections over 12 scans). In AKARI-MUSL, that condition is relaxed, and all sources detected twice or more at the same position are included regardless of how many scans pass by them (e.g., 3 detections over 12 scans). It is crucial for the Planet Nine search that the object should not be observed at the same position after half a year due to Planet Nine's parallax motion and orbital motion (see Section \ref{subsec:orb_motion} for further details). Therefore, we remove sources detected at the same position in scans separated by 6 months from the list, as those sources are probably celestial (not moving) objects.
Figure \ref{fig:detection_limit} shows a so-called $\log N$ -- $\log S$ plot, where $S$ in Jy is the source flux and $N$ is the number of objects at the flux of the most sensitive band (WIDE-S; 90 $\mu$m). AKARI-BSC's detection limits are estimated from this plot as 0.55 and 0.44 Jy for Version 1 and 2, respectively \citep{bsc_ver2}. Because sources in AKARI-MUSL are selected with a different (more relaxed) condition, it contains fainter sources down to $\sim$ 0.21 Jy. The total number of sources in MUSL, that are detected in WIDE-S, is 956,094. However, the list possibly includes many false detections caused by cosmic ray hits or instrumental artifacts. So we must carefully select possible Planet Nine candidates as we describe in the following sections.
\begin{table*}
        \centering
	\caption[Info]{Basic information of four IRAS catalogues and AKARI-MUSL. Total numbers of sources are available in published catalogues. Identified and unidentified sources of IRAS catalogues are classified by the number of positional associations ($nid$).}
	\label{tab:data}
	\begin{tabular}{ccccccc}
		\hline
        \hline
		Catalogue & Year & Wavelengths ($\mu$m) & Detection Limit (Jy) & Total Sources & Identified Sources & Unidentified Sources\\
		\hline
            IRAS-PSC & 1983 & 12, 25, 60, 100 & 0.6 & 245,889 & 85,047 & 160,842\\
            IRAS-FSC & 1983 & 12, 25, 60, 100 & 0.2 & 173,044 & 115,273 & 57,771\\
            IRAS-PSCR & 1983 & 12, 25, 60, 100 & 0.6 & 372,753 & 51,000 & 321,753\\
            IRAS-FSCR & 1983 & 12, 25, 60, 100 & 0.2 & 593,516 & 211,144 & 382,372\\
            AKARI-MUSL & 2006 & 65, 90, 140, 160 & 0.21 & 996,342 & - & -\\
            \hline
            \hline
        \end{tabular}
\end{table*}
\section{Methods}
\label{sec:method}
\subsection{Expected Orbital Motion}
\label{subsec:orb_motion}
Since the epochs of IRAS and \textit{AKARI} are separated by 23 years, it leads to the domination of orbital motion compared to the parallax motion. We therefore need to calculate the expected orbital motion of Planet Nine over 23 years in order to constrain the possible position of an IRAS candidate in the \textit{AKARI} source list. We adapted Equation (4) from \citet{Cowan2016} and rewrote it for the case of a 23-year separation as shown in Equation (\ref{eq:eq3}).
\begin{equation}
    \mu_{\mathrm{orb}} = 42\ \mathrm{arcmin}\frac{\Delta T}{23\ \mathrm{year}}\left(\frac{d}{700\ \mathrm{AU}}\right)^{-3/2}
	\label{eq:eq3}
\end{equation}
where $\mu_{\mathrm{orb}}$ is the orbital motion of Planet Nine; $\Delta T$ is the epoch separation between two surveys; $d$ is the current heliocentric distance of Planet Nine, which can be approximately equal to the geocentric distance.

When choosing an arbitrary distance range to search for Planet Nine, we can infer the angular separation between the IRAS detection and the \textit{AKARI} detection of Planet Nine over 23 years. Note that with our assumptions, Planet Nine is treated as a slow-moving object, which only moves a few arcminutes per year. It means that Planet Nine is required to have detections at different hours on the same date (hourly confirmation), and no detection on the date of 6 months before or 6 months after. In addition to the orbital motion, Planet Nine's parallax shifts every 6 months due to the Earth's motion around the Sun. As a result, we expect that Planet Nine is able to be detected in the AKARI-MUSL without monthly confirmation over 23 years. The hourly confirmation is a crucial criterion to differentiate between a slow-moving object and a fast-moving object. If an object fails to have both hourly and monthly confirmations, it is most likely a fast-moving object and cannot be considered as a good Planet Nine candidate.

\begin{figure}
   \centering
   \includegraphics[width=\columnwidth]{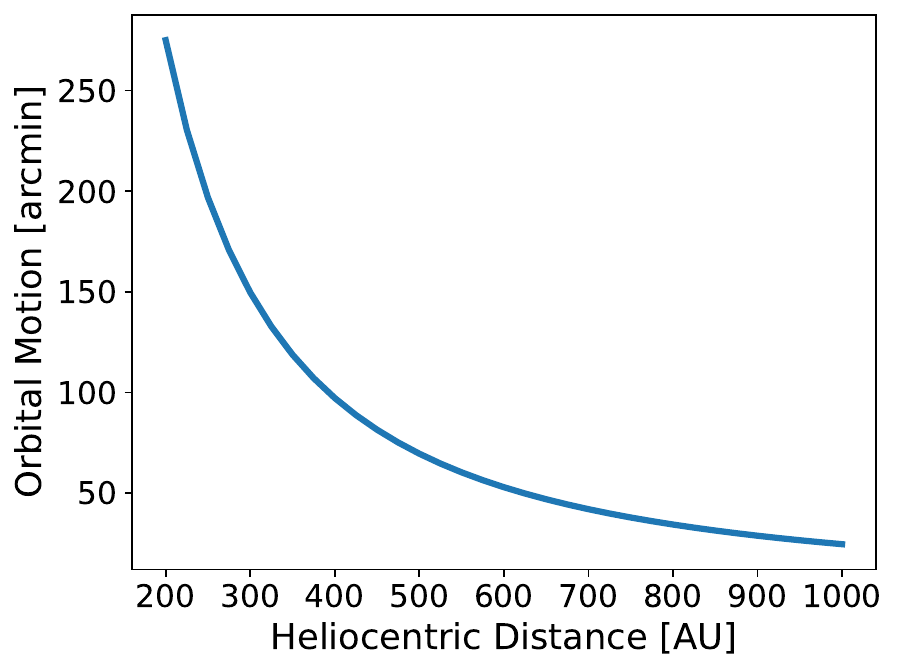}
      \caption{Expected orbital motion of Planet Nine over 23 years versus its current heliocentric distance. The orbital motion decreases exponentially as the heliocentric distance increases according to Equation (\ref{eq:eq3}).
              }
         \label{fig:orb_motion}
\end{figure}

The expected orbital motion of Planet Nine as a function of heliocentric distance is visualised in Figure \ref{fig:orb_motion}. We see that the orbital motion increases faster at shorter distances. Such a large orbital motion raises difficulties in our candidate pair selection, which are described in Section \ref{subsec:selection} and discussed in Section \ref{sec:discussion}. Therefore, in this work, we focus on the distance $\geq$ 500 AU, while the search with the distance < 500 AU will be conducted separately in our follow-up study.
\subsection{Expected Flux}
\label{subsec:flux}
Estimating the expected flux of Planet Nine is an important step to ensure that its flux must be above the detection limits of both IRAS and \textit{AKARI}. Due to our insufficient understanding of the characteristics of Planet Nine, such as mass, density, distance, temperature, etc., we need to conduct the estimation with assumptions based on simulation results from previous works. The flux density at a certain wavelength is determined by the product of two components, which are the black-body spectral radiance at that wavelength and the solid angle as seen from the Earth. Since Planet Nine was predicted to be located distantly beyond Neptune's orbit with an orbital eccentricity of 0.6 and a semi-major axis of 700 AU \citep{Batygin2016}, we expect that it is an ice-giant planet like Uranus and Neptune rather than a terrestrial planet like other inner-most planets. As a result, we assume that the density of Planet Nine is 1.46 $g/cm^{3}$, which is approximately equal to the average of the densities of Uranus (1.27 $g/cm^{3}$) and Neptune (1.64 $g/cm^{3}$), then derive the radius of Planet Nine and the solid angle observed from the Earth in steradian ($sr$). Combining the predicted semi-major axis and orbital eccentricity, we can derive the perihelion and aphelion distances of Planet Nine, which are 280 AU and 1120 AU, respectively.

To calculate the spectral radiance, we assume the black-body radiation for Planet Nine. Input parameters include frequency range ($THz$), temperature ($K$), and emissivity. Note that the sum of emissivity and Bond albedo is equal to 1. \citet{Cowan2016} assumed that Planet Nine has a similar Bond albedo to Uranus and Neptune, which are 0.30 and 0.29, respectively. Therefore, we adapted the value of 0.3 for our calculation. The spectral radiance ($W\cdot m^{-2}\cdot sr^{-1}\cdot THz^{-1}$) with such input parameters reaches the maximum value at $\sim$ 102 $\mu$m ($\sim$ 2.94 $THz$). The details of the calculation can be mathematically described in Equation (\ref{eq:eq1}) and Equation (\ref{eq:eq2}). Then we do the unit conversion from $W\cdot m^{-2}\cdot Hz^{-1}$ to Jy by using the relation 1 Jy = 10$^{-26}$ $W\cdot m^{-2}\cdot Hz^{-1}$.
\begin{equation}
    \Omega = 2\pi\left(1 - \frac{\sqrt{d^{2} - R^{2}}}{d}\right)
	\label{eq:eq1}
\end{equation}
\begin{equation}
    S_{\lambda} = L_{\lambda} \times \Omega,
	\label{eq:eq2}
\end{equation}
where $\Omega$ is the solid angle of Planet Nine observed from the Earth; $d$ is the current heliocentric distance of Planet Nine, which can be approximately equal to the geocentric distance; $R$ is the radius of Planet Nine; $S_{\lambda}$ is flux density at wavelength $\lambda$; $L_{\lambda}$ is spectral radiance at wavelength $\lambda$.

\begin{figure*}
   \centering
   \includegraphics[width=.45\columnwidth]{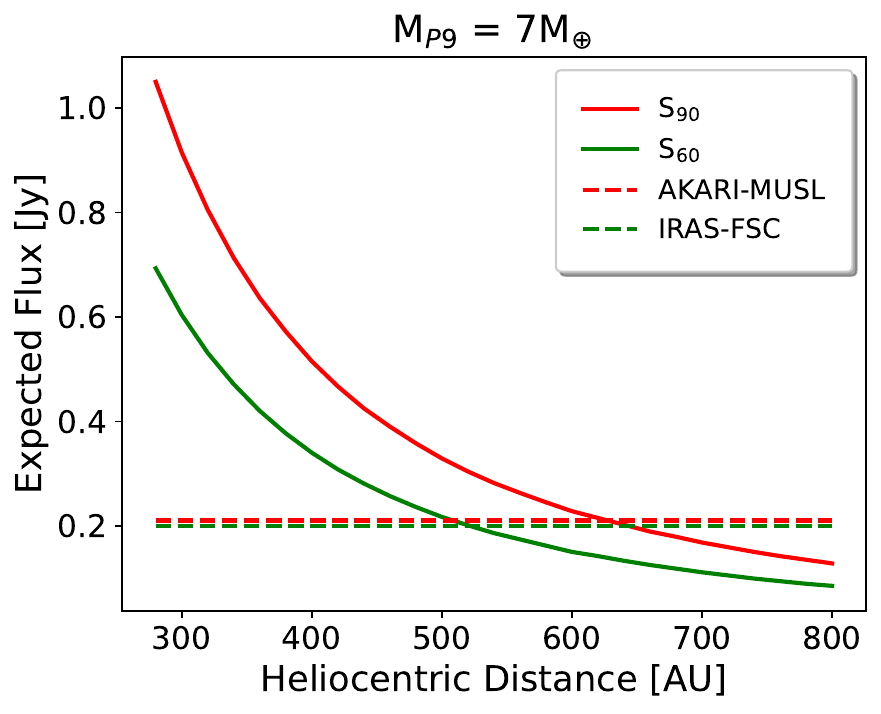}
   \includegraphics[width=.462\columnwidth]{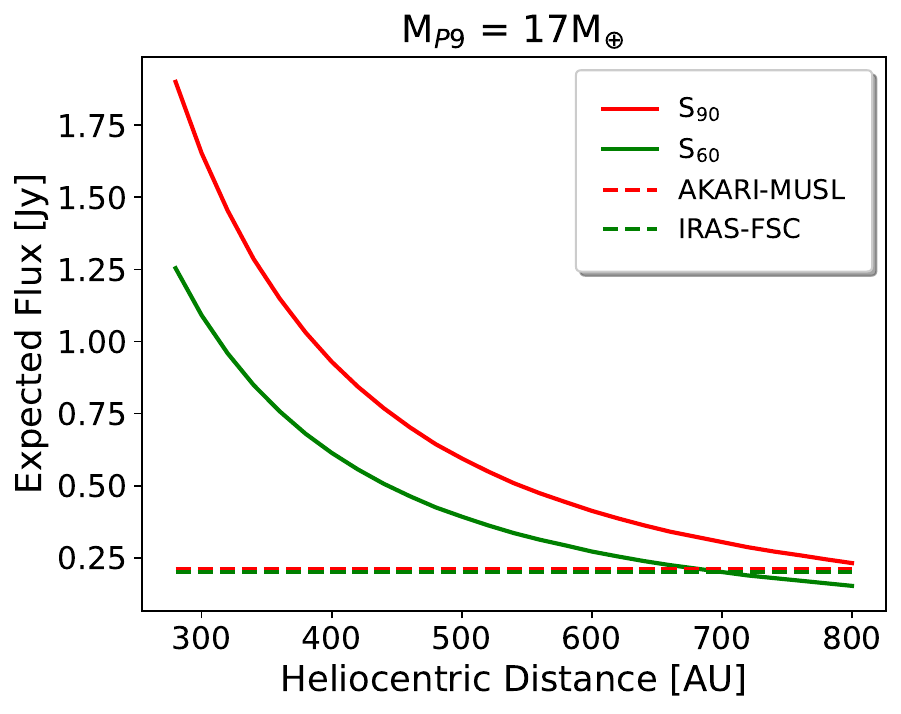}
      \caption{Expected flux of Planet Nine at 60 $\mu$m and 90 $\mu$m as a function of heliocentric distance (solid curves) compared to the flux detection limits of IRAS-FSC and AKARI-MUSL (dash lines), which are 0.2 Jy and 0.21 Jy, respectively. The expected flux is calculated with the lower limit (left panel) and the upper limit (right panel) of Planet Nine's mass range.
              }
         \label{fig:flux}
\end{figure*}

We are searching for Planet Nine candidates whose masses are less than or equal to Neptune's mass ($\sim$ 17$M_{\oplus}$) since the maximum predicted radius of Planet Nine is approximately equal to the radius of Neptune \citep{BATYGIN20191} and its assumed density is also similar to Uranus and Neptune. To determine the lower limit of the mass range, we gradually reduce the value of Planet Nine's mass to guarantee that the expected flux of Planet Nine is above the detection limits of two surveys. Figure \ref{fig:flux} shows the comparison between the expected flux of Planet Nine and the detection limits of IRAS-FSC and AKARI-MUSL at their most sensitive bands. Here we choose the detection limits of IRAS-FSC because it is three times deeper than that of IRAS-PSC at 60 $\mu$m. It is also reasonable to determine the minimum condition for Planet Nine's detection. At the upper limit of the mass range, Planet Nine would be well-detected within the heliocentric distance range of 500 -- 700 AU. However, the detectability of Planet Nine would be decreased when its mass is assumed to be lower. We notice that Planet Nine with a mass of 7$M_{\oplus}$ is detectable in IRAS within 500 -- 520 AU.
\subsection{Selection Process}
\label{subsec:selection}
We aim to search for possible Planet Nine candidate pairs, including one IRAS detection and one \textit{AKARI} detection. Criteria in our selection process can be divided into two categories, which are positional criteria and flux criteria. The order of those criteria presented in this study can be managed to ensure we obtain the same finalist. If sources match one of the following criteria, they are removed from the selection:
\begin{itemize}
    \item \textbf{IRAS or \textit{AKARI} sources with low flux quality:} Flux quality of sources is indicated by an integer number ranging from 1 to 3 (a value of 3 represents the highest flux quality). Sources whose flux quality value equals 1 at one of the following bands: 60, 65, 90, and 100 $\mu$m are referred to as unreliable flux measurements or unconfirmed sources. As a result, it is necessary to remove those sources before comparing their fluxes.\\
    \item \textbf{Below detection limits and identified sources:} Detection limits corresponding to each catalogue are provided in Table \ref{tab:data}. Already known IRAS sources are defined by the number of positional associations $nid$ = 0.\\
    \item \textbf{Too close to Galactic plane and Galactic bulge:} Sources located near the Galactic plane or the Galactic bulge could have a potential problem. For example, the brightness of the galactic cirrus may significantly affect the fluxes of those sources. Consequently, the reliability of flux measurements cannot be guaranteed. We excluded sources with low galactic latitude ($\left|b\right|$ < 10$^{\circ}$) and within a radius of 27.5$^{\circ}$ from the Galactic centre. These positional criteria come from \citet{Bernard_1994}, which showed the dominance of dust emission near the low-latitude and inner galactic regions. Similarly, Figure 3 in SS22 also indicated that without the above filters, a large number of their candidates concentrate near the Galactic plane and the Galactic centre. If Planet Nine were located in these regions, we would unfortunately fail to detect it in this study. However, the most promising region predicted by the simulation \citep{Millholland2017} is not in this direction.\\
    \item \textbf{IRAS or \textit{AKARI} fluxes exceed the upper limit of the expected flux:} This criterion helps to exclude sources with inappropriate fluxes to be Planet Nine in our search range. The upper limit was calculated by the flux estimation in Section \ref{subsec:flux} and considered as the flux of Planet Nine corresponding to the heliocentric distance of 500 AU.\\
    \item \textbf{IRAS or \textit{AKARI} flux ratios are too large:} The expected range of Planet Nine's temperature is 28 -- 53 K \citep{Cowan2016}. We exclude sources showing high temperatures in far-infrared wavelengths if the flux ratios of two bands in IRAS or \textit{AKARI} are larger than 3, corresponding to the temperature $\gtrsim$ 100 K. To be specific, $S_{100}/S_{60}$ > 3 or $S_{60}/S_{100}$ > 3 in IRAS; $S_{90}/S_{65}$ > 3 or $S_{65}/S_{90}$ > 3 in \textit{AKARI}.
\end{itemize}

In the next step, our goal is to find candidate pairs based on the expected orbital motion corresponding to the heliocentric distance range of 500 -- 700 AU. In other words, for each IRAS source, we try to search for all possible \textit{AKARI} sources with angular separation between $42'$ and $69.6'$. The angular separation increases with closer heliocentric distances to Planet Nine, which means one IRAS source could pair up with more \textit{AKARI} sources.

\begin{figure*}
   \centering
   \includegraphics[width=.627\columnwidth]{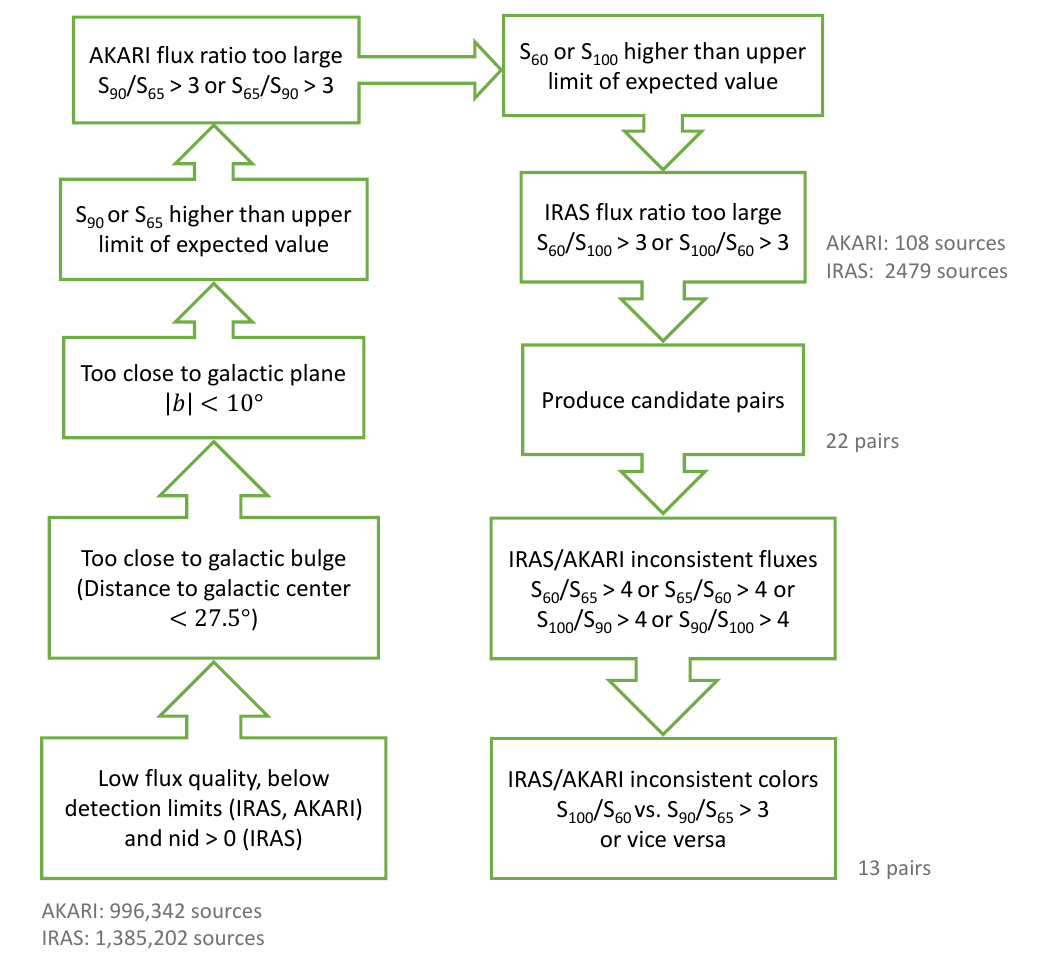}
      \caption{Flowchart of Planet Nine candidate selection process. The entire process consists of positional and flux selection criteria. Single sources or candidate pairs are excluded if they match one of the criteria.
              }
         \label{fig:flowchart}
\end{figure*}

Although these IRAS -- \textit{AKARI} pairs meet the requirement of angular separation, there might be inconsistencies in flux or colour between IRAS and \textit{AKARI} components \citep{Sedgwick2022}. We remove sources with flux inconsistency at two close wavelengths by a factor of more than 4\footnote{See Section 4.4 (b) in SS22 for the criterion of the flux inconsistency.}, i.e., sources with $S_{60}/S_{65}$ > 4; $S_{65}/S_{60}$ > 4; $S_{90}/S_{100}$ > 4; or $S_{100}/S_{90}$ > 4 are removed. 
We further require the IRAS color ($S_{100}/S_{60}$) and \textit{AKARI} color ($S_{90}/S_{65}$) to be consistent, i.e., when the color values differ by a factor of more than 3\footnote{See Section 4.4 (e) in SS22 for the criterion of the colour inconsistency.}, the sources are removed. 
The flowchart of our selection process is summarised in Figure \ref{fig:flowchart}.
\section{Results}
\label{sec:results}
After excluding IRAS -- \textit{AKARI} pairs with strongly inconsistent colours or fluxes, we found 13 candidate pairs in total, whose IRAS sources are all from IRAS-FSC.

\subsection{Image Inspection}
\label{subsec:image}
To evaluate the quality of these pairs, we implement the image inspection for both IRAS and \textit{AKARI} sources. IRAS images are downloaded from Improved Reprocessing of the IRAS Survey (IRIS) Data Access\footnote{\url{https://irsa.ipac.caltech.edu/data/IRIS/}}, while \textit{AKARI} images are downloaded from AKARI Far-Infrared All-Sky Survey Maps Data Access\footnote{\url{https://irsa.ipac.caltech.edu/data/AKARI/}}. The requirements for a good candidate pair include:
\begin{itemize}
    \item The presence of the solid IRAS source at the given coordinate in the IRAS image.\\
    \item The absence of that IRAS source at the same coordinate in the AKARI image.\\
    \item The absence of the \textit{AKARI} source at the same coordinate in the IRAS image.\\
    \item Both IRAS and \textit{AKARI} images were not contaminated by the cirrus.
\end{itemize}
If a pair fails to satisfy one of the above conditions, it is removed from the candidate list. Only one candidate pair remains.

\begin{figure*}
   \centering
   \includegraphics[width=\columnwidth]{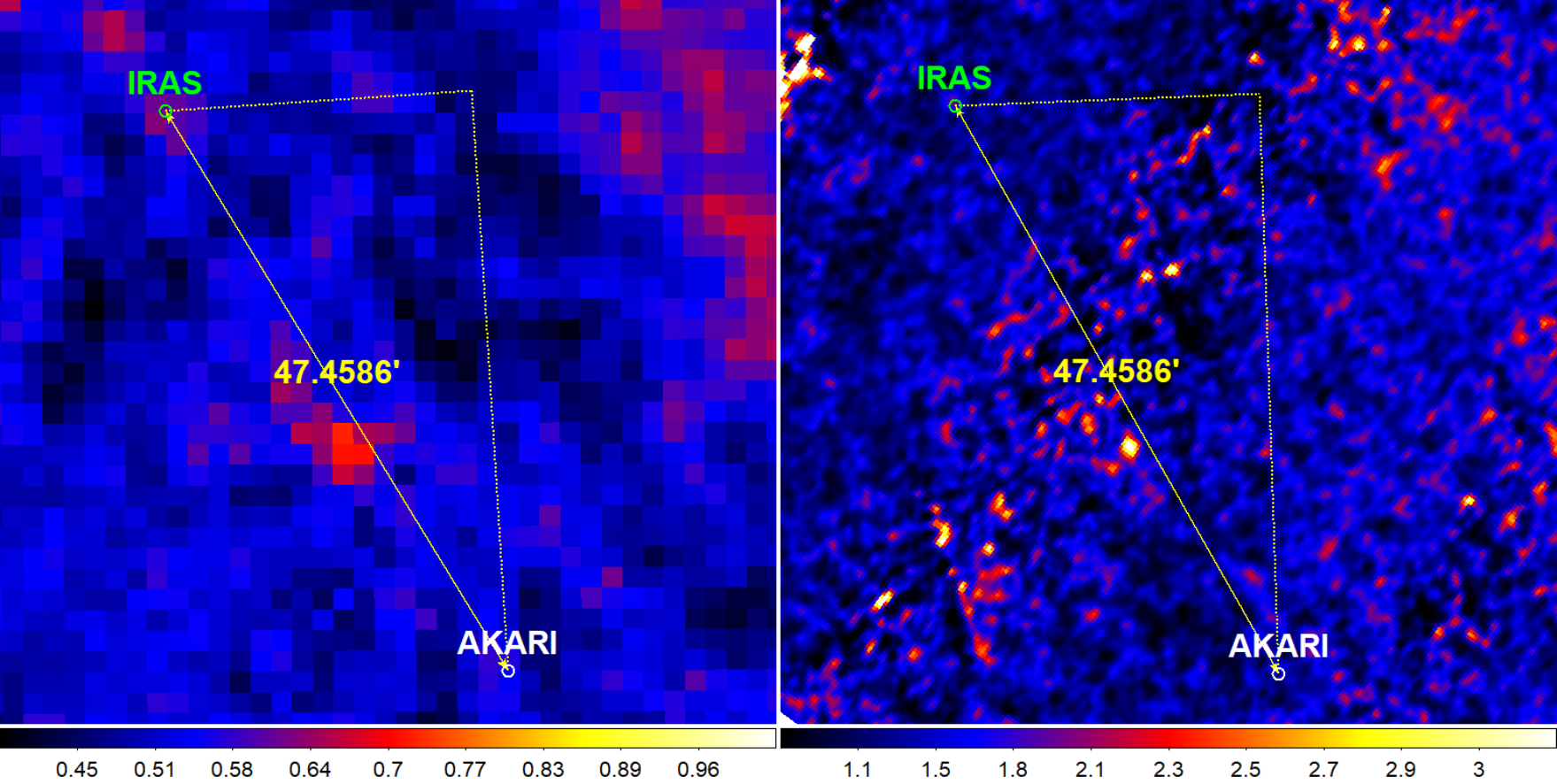}
      \caption{Comparison between IRAS (left) and \textit{AKARI} (right) cutout images of our good candidate pair. The green circle indicates the location of IRAS source, while the white circle indicates the location of \textit{AKARI} source. The size of each circle is \SI{25}{\arcsecond}. The yellow arrow with a number in arcminute shows the angular separation between IRAS and \textit{AKARI} sources. The colour bar represents the pixel intensity in each image in the unit of MJy/sr. The \textit{AKARI} source in the right panel is not visible as a real physical source due to the characteristics of AKARI-MUSL, which include moving sources without monthly confirmation.
              }
         \label{fig:pair12}
\end{figure*}

Figure \ref{fig:pair12} is our good candidate pair satisfying given conditions. The comparison between IRAS and \textit{AKARI} images indicates that this object moved out of the IRAS coordinate over 23 years. The IRAS source pairs up with only one possible source from AKARI-MUSL. The IRAS image (left panel) was taken by co-adding multiple scans, which were carried out from 26 June 1983 to 01 September 1983\footnote{We checked the date of each scan by using IRAS Scan Processing and Integration tool (SCANPI): \url{https://irsa.ipac.caltech.edu/applications/Scanpi/}}. This two-month separation corresponds to Planet Nine's orbital motion of $\sim$ \SI{18}{\arcsecond}\footnote{See the angular resolution of IRAS at \url{https://lambda.gsfc.nasa.gov/product/iras/}}, which is reasonable to be detected as a single source rather than two different sources.

We note that the \textit{AKARI} source in this pair is not visible in the co-added \textit{AKARI} image on the right panel in Figure \ref{fig:pair12}. This is due to the characteristics of AKARI-MUSL, which include moving sources without monthly confirmation. As a result, they are sometimes not visible in the co-added images, which include scans where they are not detected. The information regarding source names, coordinates, and flux densities of our good candidate pair are summarised in Table \ref{tab:good_candidate}. Although our candidate pair passed all selection criteria listed in Section \ref{subsec:selection}, $S_{65}$ is significantly fainter than $S_{60}$ and $S_{90}$. The reason for this situation is that there are still confirmed sources with very low flux compared to the detection limits in the \textit{AKARI} source list\footnote{See Section 5.1 and 7.6 in AKARI/FIS BSC Release Note Version 1 (\url{https://data.darts.isas.jaxa.jp/pub/akari/AKARI-FIS_Catalogue_AllSky_BrightSource_1.0/AKARI-FIS_BSC_V1_RN.pdf})}. Therefore, the source is confirmed at 65 $\mu$m, but the measured flux might be unreliable. It is essential to conduct follow-up observations to measure the flux of the target in other bands.
\begin{table}
	\centering
	\caption{Basic information of the good candidate pair found in this work.}
	\label{tab:good_candidate}
	\begin{tabular}{c|c}
		\hline
            \hline
            IRAS Name & F02211-4844\\
            $RA_{IRAS}$ & 35.74075$^{\circ}$\\
            $DEC_{IRAS}$ & -48.5125$^{\circ}$\\
            \hline
            \textit{AKARI} Name & AKARI-MUSL J0220440-491247\\
            $RA_{AKARI}$ & 35.18379$^{\circ}$\\
            $DEC_{AKARI}$ & -49.2135$^{\circ}$\\
            \hline
            Angular Separation & 47.4586$'$\\
            $S_{60}$ & 0.24 Jy\\
            $S_{65}$ & 0.09 Jy\\
            $S_{90}$ & 0.27 Jy\\
            $S_{100}$ & 0.52 Jy\\
            \hline
            \hline
        \end{tabular}
\end{table}
\subsection{AKARI Detection Probability Map}
\label{subsec:map}
\begin{figure}
   \centering
   \includegraphics[width=\columnwidth]{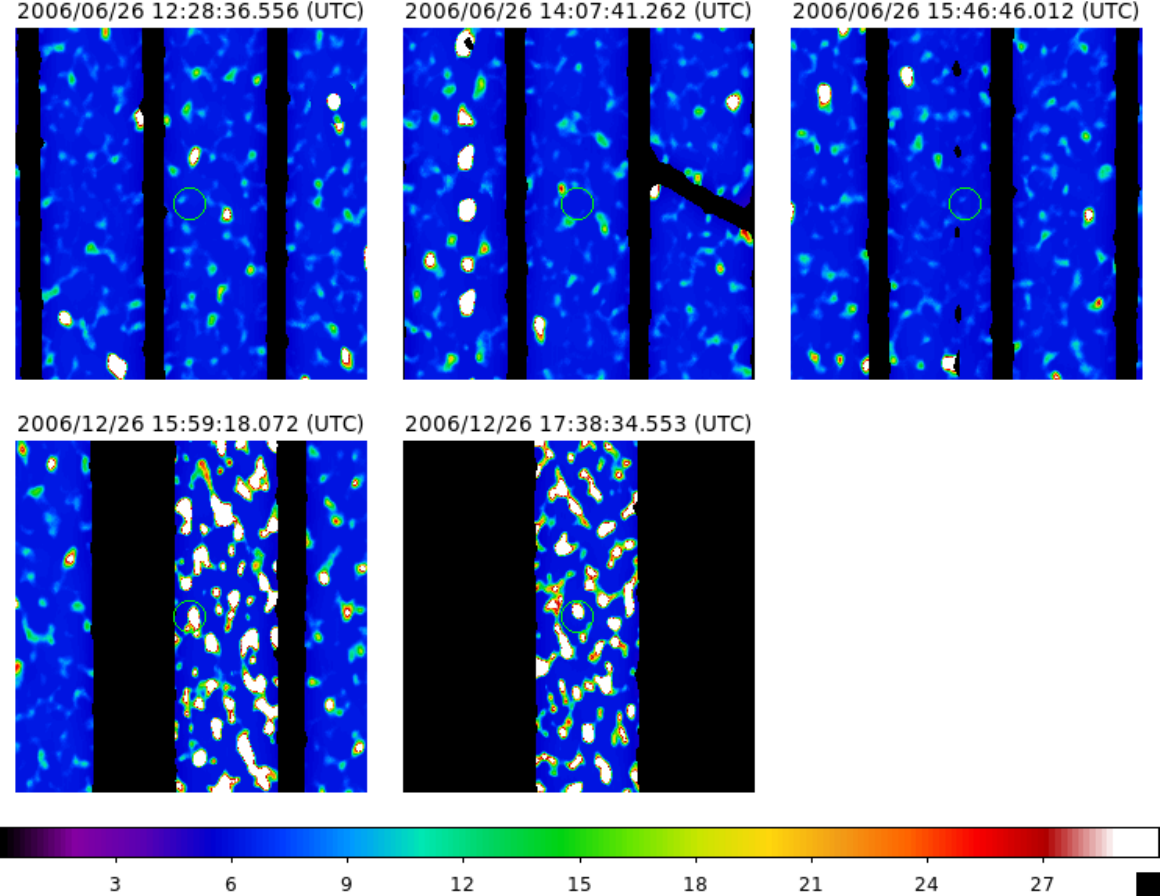}
      \caption{Detection probability map of \textit{AKARI} source in our good candidate pair. Three scans in the upper row were taken on 26 June 2006, while the other two scans in the lower row were taken on 26 December 2006. The scanning time is shown on the top of each panel along with its scanning date. The size of each panel is $30' \times 30'$. The size of the green circles at the centre of each panel is \SI{80}{\arcsecond}. The colour bar represents the pixel intensity in an arbitrary unit. In the point source extraction, pixels with intensity $\geq 15$ are treated as detections at the first step and sent to the confirmation process.
              }
         \label{fig:akarimap12}
\end{figure}

We continue to determine if the good candidate is a slow-moving or fast-moving object. This task is possible for the \textit{AKARI} component in each pair since we are able to check the probability of each detection of the \textit{AKARI} source by using the \textit{AKARI} detection probability map. Figure \ref{fig:akarimap12} shows the detection probability map of the \textit{AKARI} component in our candidate pair. \textit{AKARI} has scanned the position 5 times. Three were detected on 26 June 2006, and the other two were detected on 26 December 2006. 

A slow-moving object requires detections for all successful scans on one date and no detection on the other date, which means that the \textit{AKARI} source does not have monthly confirmation. To be specific, the expected angular separation of Planet Nine over 6 months at 700 AU is roughly \SI{55}{\arcsecond}, which is distinguishable and should be considered as two different detections. On the date of detection, there must be at least two detections at different times, which means that the \textit{AKARI} source has hourly confirmation. An additional criterion for good detection is that the green circles are required not to reach the edge of each scan. In other words, the green circles are completely located in the blue region. In this case, there is almost no detection in the three upper panels on the same date, while we obviously see two detections after 6 months in the two lower panels. Therefore, the \textit{AKARI} source in our candidate pair is most likely a slow-moving object.
\section{Discussions}
\label{sec:discussion}
The finalist of our Planet Nine candidate pair strongly depends on how the characteristics of Planet Nine are defined. Indeed, if the actual mass of Planet Nine is larger than that of Neptune, Planet Nine still has a possibility to be detected in IRAS and \textit{AKARI} due to its higher flux in far-infrared wavelengths. In contrast, if the actual mass of Planet Nine is not sufficient to make its flux above the detection limits of two surveys, there is no chance of finding Planet Nine in this work. Similarly, the heliocentric distance range of Planet Nine significantly affects the total number of candidate pairs found by our algorithm. For example, the angular separation between IRAS and \textit{AKARI} sources after 23 years becomes larger when the lower limit of the heliocentric distance range is smaller. As a result, the pair-searching radius and the total number of our candidate pairs will be increased. In addition, our understanding of the effective temperature and the density of Planet Nine has not been completed to date, therefore, the assumptions for these parameters are inevitable and rely on the results of previous simulation works.

Obvious evidence for the dependence of Planet Nine candidates on the assumed parameters is the results reported in SS22. Their study covered the distance range from 700 AU to 8000 AU, which is further and wider than ours. Although they also searched for Planet Nine candidate pairs using IRAS and \textit{AKARI}, the IRAS components in most of their candidate pairs are from IRAS-PSCR, while in our study, the IRAS components are all from IRAS-FSC. There are various explanations for the difference between the two results. First, SS22 used different assumptions and models from our work to estimate the expected flux and orbital motion of Planet Nine. Table 4 in SS22 indicates that they searched for brighter giant planets with smaller angular separation between IRAS and \textit{AKARI} sources. Second, SS22 did not exclude sources below the flux detection limits of IRAS and \textit{AKARI}. Again, according to Table 4 in SS22, Uranus-mass planets with a distance further than 2000 AU have $S_{60} \leq$ 0.07 Jy and $S_{90} \leq$ 0.09 Jy, which are all below the flux detection limits of IRAS and AKARI-BSC Version 2. Third, before comparing fluxes between different wavelengths or to the flux detection limits, we excluded IRAS and \textit{AKARI} sources with low flux quality, while SS22 did not. This was helpful in avoiding the unreliable sources in our selection.

In comparison to the previous search for Planet Nine using IRAS data \citep{Rowan2021}, there are some discrepancies regarding the strategy and the outcome. First, \citet{Rowan2021} also searched for unidentified slow-moving objects without months confirmation, but only the IRAS survey was utilised for the Planet Nine search. Second, the outcome of their study proposed one candidate, which is in IRAS-PSCR with single hour confirmation. Interestingly, the fitted orbit indicated that their candidate has a distance of 225 $\pm$ 15 AU, which is even closer than the distance range predicted in \citet{Batygin2016}. In addition, the mass range of 3 -- 5M$_{\oplus}$ is also lower than that of this work. This could be explained by the different expectations of the characteristics of Planet Nine between \citet{Rowan2021} and this work. To be specific, we expected that the flux ratio between 60 $\mu$m and 100 $\mu$m bands should be no more than 3, while that ratio is approximately 5 as shown in Table 1 of \citet{Rowan2021}. They also concluded that their proposed candidate, if real, may not cause the orbital clustering of minor KBOs due to the difference in direction on the sky compared to \citet{Brown2021}.

Although we took advantage of the \textit{AKARI} detection probability map of the \textit{AKARI} source to determine whether it is a fast-moving or slow-moving object, two detections from IRAS and \textit{AKARI} are not sufficient to determine the Keplerian motion of our candidate. In addition, the \textit{AKARI} source is confirmed at 65 $\mu$m, but its flux measurement might be unreliable. It is essential to figure out solutions for these issues. The follow-up observation using Dark Energy Camera (DECam) Imager mounted on the Blanco telescope with 3 deg$^{2}$ field of view\footnote{See the detail of Blanco telescope's horizon limits at \url{https://noirlab.edu/science/programs/ctio/telescopes/victor-blanco-4m-telescope/Horizon-Limits}} can offer a promising opportunity to examine the Keplerian motion of our good candidate pair. Figure 9 in \citet{Brown2021} indicates that the predicted r-band magnitude of Planet Nine is less than 26. Such an upper limit of 26 AB mag only needs roughly one hour to reach a signal-to-noise ratio of 5 using the r-band filter of DECam\footnote{The DECam Exposure Time Calculator Ver 7B is available to download at \url{https://noirlab.edu/science/documents/scidoc0493}}. \citet{Brown2024} pointed out that 9 detections are required to evaluate a linked Keplerian orbit with improved processing speed and reduce the number of false positives. The method of the follow-up analysis should be similar to this study. We can start to estimate the expected orbital motion over 18 years (from 2006 to 2024), which varies between $33'$ and $54.7'$, corresponding to the heliocentric distance range of 500 -- 700 AU. If the existence of Planet Nine can be confirmed by observations in the near future, it will improve our understanding of the history and structure of the entire solar system in early stages.
\section{Summary}
\label{sec:sum}
We searched for Planet Nine candidates in a heliocentric distance range of 500 -- 700 AU and a mass range of 7 -- 17M$_{\oplus}$ by using two far-infrared all-sky surveys with a 23-year epoch difference. Planet Nine is expected to move slowly on the sky due to its great distance beyond Neptune's orbit. Therefore, we searched for slow-moving objects that moved from an IRAS position to another \textit{AKARI} position after 23 years. The expected flux of Planet Nine was estimated by assuming the black-body radiation in infrared wavelengths. The outcomes of this research are summarised as follows:
\begin{itemize}
    \item After the rigorous selection including the visual image inspection, we found one good candidate pair, in which the IRAS source was not detected at the same position in the \textit{AKARI} image and vice versa, with the expected angular separation of $42'$ -- $69.6'$.\\
    \item The \textit{AKARI} detection probability map indicated that the \textit{AKARI} source of our candidate pair satisfied the requirements for a slow-moving object with two detections on one date and no detection on the date of 6 months before.
\end{itemize}
However, the IRAS and \textit{AKARI} detections alone are not enough to decide a precise orbit. The DECam, with a large field of view, is a prospective option for the follow-up observation. It enables the possibility of detecting faint moving objects even in optical wavelengths and determining the full orbit of our candidate, since the exposure time to observe targets as faint as 26 AB mag at DECam is approximately one hour. The verification of Planet Nine's existence via future observational studies will contribute to our understanding of the evolution and structural dynamics of the solar system.

\begin{acknowledgement}
We would like to thank the anonymous reviewers for their careful reading and constructive comments, which significantly helped to improve the quality of the manuscript. TG acknowledges the support of the National Science and Technology Council of Taiwan (NSTC) through grants 108-2628-M-007-004-MY3, 110-2112-M-005-013-MY3, 111-2112-M-007-021, 111-2123-M-001-008-, 112-2112-M-007-013, 112-2123-M-001-004-, 113-2112-M-007-006-, 113-2927-I-007-501-, and 113-2123-M-001-008-.
TH acknowledges the support of the National Science and Technology Council of Taiwan through grants 110-2112-M-005-013-MY3, 110-2112-M-007-034-, and 113-2123-M-001-008-.
TN acknowledges the support by the JSPS Kakenhi grants 23H05441 and 23K17695.
SH acknowledges the support of the Australian Research Council (ARC) Centre of Excellence (CoE) for Gravitational Wave Discovery (OzGrav) project numbers CE170100004 and CE230100016, and the ARC CoE for All Sky Astrophysics in 3 Dimensions (ASTRO 3D) project number CE170100013.
This research is based on observations with \textit{AKARI}, a JAXA project with the participation of ESA.
The Infrared Astronomical Satellite (IRAS) was a joint project of the US, UK and the Netherlands.
This research has made use of the NASA/IPAC Infrared Science Archive, which is funded by the National Aeronautics and Space Administration and operated by the California Institute of Technology.
The authors are thankful to our collaborators from NTHU \& NCHU Cosmology Group, including Shotaro Yamasaki, Ece Kilerci, and Murthadza Aznam, for their insightful discussions.
\end{acknowledgement}

\section*{Data Availability Statement}
The datasets of four IRAS catalogues are available at \url{https://irsa.ipac.caltech.edu/Missions/iras.html}.
The dataset of AKARI-MUSL will be made available upon reasonable request to the corresponding author.

\printendnotes

\printbibliography

@article{Batygin2016,
url = {https://dx.doi.org/10.3847/0004-6256/151/2/22},
year = {2016},
month = {1},
publisher = {The American Astronomical Society},
volume = {151},
number = {2},
pages = {22},
author = {Konstantin Batygin and Michael E. Brown},
title = {EVIDENCE FOR A DISTANT GIANT PLANET IN THE SOLAR SYSTEM},
journal = {The Astronomical Journal}
}

@article{Brown2019,
url = {https://dx.doi.org/10.3847/1538-3881/aaf051},
year = {2019},
month = {1},
publisher = {The American Astronomical Society},
volume = {157},
number = {2},
pages = {62},
author = {Michael E. Brown and Konstantin Batygin},
title = {Orbital Clustering in the Distant Solar System},
journal = {The Astronomical Journal}
}

@article{Brown2021,
title = {The orbit of Planet Nine},
author = {Michael E. Brown and Konstantin Batygin},
url = {https://dx.doi.org/10.3847/1538-3881/ac2056},
year = {2021},
month = {10},
publisher = {The American Astronomical Society},
volume = {162},
number = {5},
pages = {219},
journal = {The Astronomical Journal}
}

@article{trujillo2014sedna,
  title={A Sedna-like body with a perihelion of 80 astronomical units},
  author={Trujillo, Chadwick A and Sheppard, Scott S},
  journal={Nature},
  volume={507},
  number={7493},
  pages={471--474},
  year={2014},
  publisher={Nature Publishing Group UK London},
  url = {https://doi.org/10.1038/nature13156}
}

@article{Brown2022,
url = {https://dx.doi.org/10.3847/1538-3881/ac32dd},
year = {2022},
month = {1},
publisher = {The American Astronomical Society},
volume = {163},
number = {2},
pages = {102},
author = {Michael E. Brown and Konstantin Batygin},
title = {A Search for Planet Nine using the Zwicky Transient Facility Public Archive},
journal = {The Astronomical Journal}
}

@article{Cowan2016,
url = {https://dx.doi.org/10.3847/2041-8205/822/1/L2},
year = {2016},
month = {4},
publisher = {The American Astronomical Society},
volume = {822},
number = {1},
pages = {L2},
author = {Nicolas B. Cowan and Gil Holder and Nathan A. Kaib},
title = {COSMOLOGISTS IN SEARCH OF PLANET NINE: THE CASE FOR CMB EXPERIMENTS},
journal = {The Astrophysical Journal Letters}
}

@article{Rowan2021,
    author = {Rowan-Robinson, Michael},
    title = "{A search for Planet 9 in the IRAS data}",
    journal = {Monthly Notices of the Royal Astronomical Society},
    volume = {510},
    number = {3},
    pages = {3716-3726},
    year = {2021},
    month = {11},
    issn = {0035-8711},
    url = {https://doi.org/10.1093/mnras/stab3212},
}

@article{Sedgwick2022,
    author = {Sedgwick, Chris and Serjeant, Stephen},
    title = "{Searching for giant planets in the outer Solar system with far-infrared all-sky surveys}",
    journal = {Monthly Notices of the Royal Astronomical Society},
    year = {2022},
    volume = {515},
    number = {4},
    pages = {4828-4837},
    month = {7},
    issn = {0035-8711},
    url = {https://doi.org/10.1093/mnras/stac2044},
}

@article{Millholland2017,
url = {https://dx.doi.org/10.3847/1538-3881/153/3/91},
year = {2017},
month = {2},
publisher = {The American Astronomical Society},
volume = {153},
number = {3},
pages = {91},
author = {Sarah Millholland and Gregory Laughlin},
title = {Constraints on Planet Nine’s Orbit and Sky Position within a Framework of Mean-motion Resonances},
journal = {The Astronomical Journal}
}

@article{Brown2024,
url = {https://dx.doi.org/10.3847/1538-3881/ad24e9},
year = {2024},
month = {3},
publisher = {The American Astronomical Society},
volume = {167},
number = {4},
pages = {146},
author = {Michael E. Brown and Matthew J. Holman and Konstantin Batygin},
title = {A Pan-STARRS1 Search for Planet Nine},
journal = {The Astronomical Journal}
}

@article{Belyakov2022,
url = {https://dx.doi.org/10.3847/1538-3881/ac5c56},
year = {2022},
month = {4},
publisher = {The American Astronomical Society},
volume = {163},
number = {5},
pages = {216},
author = {Matthew Belyakov and Pedro H. Bernardinelli and Michael E. Brown},
title = {Limits on the Detection of Planet Nine in the Dark Energy Survey},
journal = {The Astronomical Journal}
}

@article{neugebauer1984infrared,
  title={The Infrared Astronomical Satellite (IRAS) mission},
  author={Neugebauer, G and Habing, HJ and Van Duinen, R and Aumann, HH and Baud, B and Beichman, CA and Beintema, DA and Boggess, N and Clegg, PE and De Jong, Teije and others},
  journal={The Astrophysical Journal},
  volume={278},
  pages={L1--L6},
  year={1984},
  month={3},
  url={https://ui.adsabs.harvard.edu/abs/1984ApJ...278L...1N}
}

@article{Murakami2007,
    author = {Murakami, Hiroshi and Baba, Hajime and Barthel, Peter and Clements, David L. and Cohen, Martin and Doi, Yasuo and Enya, Keigo and Figueredo, Elysandra and Fujishiro, Naofumi and Fujiwara, Hideaki and Fujiwara, Mikio and Garcia-Lario, Pedro and Goto, Tomotsugu and Hasegawa, Sunao and Hibi, Yasunori and Hirao, Takanori and Hiromoto, Norihisa and Hong, Seung Soo and Imai, Koji and Ishigaki, Miho and Ishiguro, Masateru and Ishihara, Daisuke and Ita, Yoshifusa and Jeong, Woong-Seob and Jeong, Kyung Sook and Kaneda, Hidehiro and Kataza, Hirokazu and Kawada, Mitsunobu and Kawai, Toshihide and Kawamura, Akiko and Kessler, Martin F. and Kester, Do and Kii, Tsuneo and Kim, Dong Chan and Kim, Woojung and Kobayashi, Hisato and Koo, Bon Chul and Kwon, Suk Minn and Lee, Hyung Mok and Lorente, Rosario and Makiuti, Sin’itirou and Matsuhara, Hideo and Matsumoto, Toshio and Matsuo, Hiroshi and Matsuura, Shuji and MÜller, Thomas G. and Murakami, Noriko and Nagata, Hirohisa and Nakagawa, Takao and Naoi, Takahiro and Narita, Masanao and Noda, Manabu and Oh, Sang Hoon and Ohnishi, Akira and Ohyama, Youichi and Okada, Yoko and Okuda, Haruyuki and Oliver, Sebastian and Onaka, Takashi and Ootsubo, Takafumi and Oyabu, Shinki and Pak, Soojong and Park, Yong-Sun and Pearson, Chris P. and Rowan-Robinson, Michael and Saito, Toshinobu and Sakon, Itsuki and Salama, Alberto and Sato, Shinji and Savage, Richard S. and Serjeant, Stephen and Shibai, Hiroshi and Shirahata, Mai and Sohn, Jungjoo and Suzuki, Toyoaki and Takagi, Toshinobu and Takahashi, Hidenori and TanabÉ, Toshihiko and Takeuchi, Tsutomu T. and Takita, Satoshi and Thomson, Matthew and Uemizu, Kazunori and Ueno, Munetaka and Usui, Fumihiko and Verdugo, Eva and Wada, Takehiko and Wang, Lingyu and Watabe, Toyoki and Watarai, Hidenori and White, Glenn J. and Yamamura, Issei and Yamauchi, Chisato and Yasuda, Akiko},
    title = "{The Infrared Astronomical Mission AKARI*}",
    journal = {Publications of the Astronomical Society of Japan},
    volume = {59},
    number = {sp2},
    pages = {S369-S376},
    year = {2007},
    month = {10},
    issn = {0004-6264},
    url = {https://doi.org/10.1093/pasj/59.sp2.S369},
}

@article{Brown2004,
url = {https://dx.doi.org/10.1086/422095},
year = {2004},
month = {12},
publisher = {},
volume = {617},
number = {1},
pages = {645},
author = {Michael E. Brown and Chadwick Trujillo and David Rabinowitz},
title = {Discovery of a Candidate Inner Oort Cloud Planetoid},
journal = {The Astrophysical Journal}
}

@article{PhysRevLett.125.051103,
  title = {What If Planet 9 Is a Primordial Black Hole?},
  author = {Scholtz, Jakub and Unwin, James},
  journal = {Phys. Rev. Lett.},
  volume = {125},
  issue = {5},
  pages = {051103},
  numpages = {7},
  year = {2020},
  month = {7},
  publisher = {American Physical Society},
  url = {https://link.aps.org/doi/10.1103/PhysRevLett.125.051103}
}

@article{Batygin_2024,
url = {https://dx.doi.org/10.3847/2041-8213/ad3cd2},
year = {2024},
month = {4},
publisher = {The American Astronomical Society},
volume = {966},
number = {1},
pages = {L8},
author = {Konstantin Batygin and Alessandro Morbidelli and Michael E. Brown and David Nesvorný},
title = {Generation of Low-inclination, Neptune-crossing Trans-Neptunian Objects by Planet Nine},
journal = {The Astrophysical Journal Letters}
}

@article{Sheppard_2019,
url = {https://dx.doi.org/10.3847/1538-3881/ab0895},
year = {2019},
month = {3},
publisher = {The American Astronomical Society},
volume = {157},
number = {4},
pages = {139},
author = {Scott S. Sheppard and Chadwick A. Trujillo and David J. Tholen and Nathan Kaib},
title = {A New High Perihelion Trans-Plutonian Inner Oort Cloud Object: 2015 TG387},
journal = {The Astronomical Journal}
}

@article{Brown_2017,
url = {https://dx.doi.org/10.3847/1538-3881/aa79f4},
year = {2017},
month = {7},
publisher = {The American Astronomical Society},
volume = {154},
number = {2},
pages = {65},
author = {Michael E. Brown},
title = {Observational Bias and the Clustering of Distant Eccentric Kuiper Belt Objects},
journal = {The Astronomical Journal}
}

@article{BATYGIN20191,
title = {The planet nine hypothesis},
journal = {Physics Reports},
volume = {805},
pages = {1-53},
year = {2019},
note = {The planet nine hypothesis},
issn = {0370-1573},
doi = {https://doi.org/10.1016/j.physrep.2019.01.009},
author = {Konstantin Batygin and Fred C. Adams and Michael E. Brown and Juliette C. Becker}
}

@article{Brown_2016,
url = {https://dx.doi.org/10.3847/2041-8205/824/2/L23},
year = {2016},
month = {6},
publisher = {The American Astronomical Society},
volume = {824},
number = {2},
pages = {L23},
author = {Michael E. Brown and Konstantin Batygin},
title = {OBSERVATIONAL CONSTRAINTS ON THE ORBIT AND LOCATION OF PLANET NINE IN THE OUTER SOLAR SYSTEM},
journal = {The Astrophysical Journal Letters},
}

@article{Batygin_2016,
url = {https://dx.doi.org/10.3847/2041-8205/833/1/L3},
year = {2016},
month = {11},
publisher = {The American Astronomical Society},
volume = {833},
number = {1},
pages = {L3},
author = {Konstantin Batygin and Michael E. Brown},
title = {GENERATION OF HIGHLY INCLINED TRANS-NEPTUNIAN OBJECTS BY PLANET NINE},
journal = {The Astrophysical Journal Letters},
}

@ARTICLE{Bernard_1994,
       author = {{Bernard}, J.~P. and {Boulanger}, F. and {Desert}, F.~X. and {Giard}, M. and {Helou}, G. and {Puget}, J.~L.},
        title = "{Dust emission of galactic cirrus from DIRBE observations.}",
      journal = {Astronomy and Astrophysics},
         year = {1994},
        month = {11},
       volume = {291},
        pages = {L5-L8},
       url = {https://ui.adsabs.harvard.edu/abs/1994A&A...291L...5B}
}

@article{Napier_2021,
url = {https://dx.doi.org/10.3847/PSJ/abe53e},
year = {2021},
month = {3},
publisher = {The American Astronomical Society},
volume = {2},
number = {2},
pages = {59},
author = {Napier, K. J. and Gerdes, D. W. and Lin, Hsing Wen and Hamilton, S. J. and Bernstein, G. M. and Bernardinelli, P. H. and Abbott, T. M. C. and Aguena, M. and Annis, J. and Avila, S. and Bacon, D. and Bertin, E. and Brooks, D. and Burke, D. L. and Carnero Rosell, A. and Carrasco Kind, M. and Carretero, J. and Costanzi, M. and da Costa, L. N. and De Vicente, J. and Diehl, H. T. and Doel, P. and Everett, S. and Ferrero, I. and Fosalba, P. and García-Bellido, J. and Gruen, D. and Gruendl, R. A. and Gutierrez, G. and Hollowood, D. L. and Honscheid, K. and Hoyle, B. and James, D. J. and Kent, S. and Kuehn, K. and Kuropatkin, N. and Maia, M. A. G. and Menanteau, F. and Miquel, R. and Morgan, R. and Palmese, A. and Paz-Chinchón, F. and Plazas, A. A. and Sanchez, E. and Scarpine, V. and Serrano, S. and Sevilla-Noarbe, I. and Smith, M. and Suchyta, E. and Swanson, M. E. C. and To, C. and Walker, A. R. and Wilkinson, R. D. and (DES Collaboration)},
title = {No Evidence for Orbital Clustering in the Extreme Trans-Neptunian Objects},
journal = {The Planetary Science Journal},
}

@article{Shankman_2017,
url = {https://dx.doi.org/10.3847/1538-3881/aa7aed},
year = {2017},
month = {7},
publisher = {The American Astronomical Society},
volume = {154},
number = {2},
pages = {50},
author = {Shankman, Cory and Kavelaars, J. J. and Bannister, Michele T. and Gladman, Brett J. and Lawler, Samantha M. and Chen, Ying-Tung and Jakubik, Marian and Kaib, Nathan and Alexandersen, Mike and Gwyn, Stephen D. J. and Petit, Jean-Marc and Volk, Kathryn},
title = {OSSOS. VI. Striking Biases in the Detection of Large Semimajor Axis Trans-Neptunian Objects},
journal = {The Astronomical Journal},
}

@ARTICLE{Iorio_2017,
       author = {{Iorio}, Lorenzo},
        title = "{Preliminary constraints on the location of the recently hypothesized new planet of the Solar System from planetary orbital dynamics}",
      journal = {Astrophysics and Space Science},
         year = {2017},
        month = {1},
       volume = {362},
       number = {1},
          eid = {11},
        pages = {11},
       url = {https://ui.adsabs.harvard.edu/abs/2017Ap&SS.362...11I}
}

@ARTICLE{Fienga_2020,
       author = {{Fienga}, A. and {Di Ruscio}, A. and {Bernus}, L. and {Deram}, P. and {Durante}, D. and {Laskar}, J. and {Iess}, L.},
        title = "{New constraints on the location of P9 obtained with the INPOP19a planetary ephemeris}",
      journal = {Astronomy and Astrophysics},
         year = {2020},
        month = {8},
       volume = {640},
          eid = {A6},
        pages = {A6},
       url = {https://ui.adsabs.harvard.edu/abs/2020A&A...640A...6F}
}

@article{Gomes_2023,
url = {https://dx.doi.org/10.3847/PSJ/acc7a2},
year = {2023},
month = {4},
publisher = {The American Astronomical Society},
volume = {4},
number = {4},
pages = {66},
author = {Gomes, Daniel C. H. and Murray, Zachary and Gomes, Rafael C. H. and Holman, Matthew J. and Bernstein, Gary M.},
title = {Can the Gravitational Effect of Planet X be Detected in Current-era Tracking of the Known Major and Minor Planets?},
journal = {The Planetary Science Journal},
}

@ARTICLE{Nabiyev_2022,
       author = {{Nabiyev}, Shaig and {Yalim}, Jason and {Guliyev}, Ayyub and {Guliyev}, Rustam},
        title = "{Hyperbolic comets as an indicator of a hypothetical planet 9 in the solar system}",
      journal = {Advances in Space Research},
         year = {2022},
        month = {4},
       volume = {69},
       number = {8},
        pages = {3182-3203},
       url = {https://ui.adsabs.harvard.edu/abs/2022AdSpR..69.3182N}
}

@ARTICLE{Holman_2016,
       author = {{Holman}, Matthew J. and {Payne}, Matthew J.},
        title = "{Observational Constraints on Planet Nine: Cassini Range Observations}",
      journal = {The Astronomical Journal},
         year = {2016},
        month = {10},
       volume = {152},
       number = {4},
          eid = {94},
        pages = {94},
       url = {https://ui.adsabs.harvard.edu/abs/2016AJ....152...94H}
}

@ARTICLE{Fienga_2016,
       author = {{Fienga}, A. and {Laskar}, J. and {Manche}, H. and {Gastineau}, M.},
        title = "{Constraints on the location of a possible 9th planet derived from the Cassini data}",
      journal = {Astronomy and Astrophysics},
         year = {2016},
        month = {3},
       volume = {587},
          eid = {L8},
        pages = {L8},
       url = {https://ui.adsabs.harvard.edu/abs/2016A&A...587L...8F}
}

@ARTICLE{Iorio_2013,
       author = {{Iorio}, L.},
        title = "{Perspectives on effectively constraining the location of a massive trans-Plutonian object with the New Horizons spacecraft: a sensitivity analysis}",
      journal = {Celestial Mechanics and Dynamical Astronomy},
         year = {2013},
        month = {8},
       volume = {116},
       number = {4},
        pages = {357-366},
       url = {https://ui.adsabs.harvard.edu/abs/2013CeMDA.116..357I}
}

@ARTICLE{Bucko_2023,
       author = {{Bucko}, Jozef and {Soyuer}, Deniz and {Zwick}, Lorenz},
        title = "{Prospects for localizing Planet 9 with a future Uranus mission}",
      journal = {Monthly Notices of the Royal Astronomical Society: Letters},
         year = {2023},
        month = {9},
       volume = {524},
       number = {1},
        pages = {L32-L37},
       url = {https://ui.adsabs.harvard.edu/abs/2023MNRAS.524L..32B}
}

@article{Iorio_2023,
    author = {Iorio, Lorenzo and Girija, Athul Pradeepkumar and Durante, Daniele},
    title = {One EURO for Uranus: the Elliptical Uranian Relativity Orbiter mission},
    journal = {Monthly Notices of the Royal Astronomical Society},
    volume = {523},
    number = {3},
    pages = {3595-3614},
    year = {2023},
    month = {05},
    url = {https://doi.org/10.1093/mnras/stad1446},
}

@ARTICLE{Buscaino_2015,
       author = {{Buscaino}, Brandon and {DeBra}, Daniel and {Graham}, Peter W. and {Gratta}, Giorgio and {Wiser}, Timothy D.},
        title = "{Testing long-distance modifications of gravity to 100 astronomical units}",
      journal = {Physical Review D},
         year = {2015},
        month = {11},
       volume = {92},
       number = {10},
          eid = {104048},
        pages = {104048},
       url = {https://ui.adsabs.harvard.edu/abs/2015PhRvD..92j4048B}
}

@book{Iorio_2024, 
    title={General Post-Newtonian Orbital Effects: From Earth’s Satellites to the Galactic Centre}, 
    publisher={Cambridge University Press}, 
    author={Iorio, Lorenzo}, 
    year={2024},
    url = {https://ui.adsabs.harvard.edu/abs/2024gpno.book.....I}
}

@INPROCEEDINGS{bsc_ver2,
       author = {{Yamamura}, Issei and {Makiuchi}, Shinichiro and {Koga}, Tatsuya and {AKARI Team}},
        title = "{AKARI Far-Infrared Point Source Catalogues}",
    booktitle = {The Cosmic Wheel and the Legacy of the AKARI Archive: From Galaxies and Stars to Planets and Life},
         year = {2018},
       editor = {{Ootsubo}, Takafumi and {Yamamura}, Issei and {Murata}, Kazumi and {Onaka}, Takashi},
        month = {3},
        pages = {227-230},
       url = {https://ui.adsabs.harvard.edu/abs/2018cwla.conf..227Y}
}

\end{document}